\documentclass{PoS}

\def\beq{\begin{equation}}
\def\eeq{\end{equation}}
\def\beqa{\begin{eqnarray}}
\def\eeqa{\end{eqnarray}}

\usepackage{epsfig}

\title{High-precision theory for top-quark production}

\ShortTitle{High-precision theory for top-quark production}

\author{\speaker{Nikolaos Kidonakis}\thanks{This material is based upon work supported by the National Science Foundation under Grant No. PHY 1519606.}\\
  Department of Physics, Kennesaw State University, Kennesaw, GA 30144, USA\\
        E-mail: \email{nkidonak@kennesaw.edu}}

\abstract{I present high-precision results for top quark production at hadron collider energies. Total and differential cross sections are calculated through aN$^3$LO for top-antitop pair production and through aNNLO for single-top production. Top production in association with a charged Higgs boson and via anomalous couplings is also briefly discussed.}

\FullConference{XIII International Conference on Heavy Quarks and Leptons\\
		22-27 May, 2016\\
		Blacksburg, Virginia, USA}

\begin{document}

\section{Introduction}

The top quark is the heaviest elementary particle ever discovered and thus holds a very special place in the list of fundamental particles. It is predominantly produced in hadron colliders via top-antitop pair production processes and to a lesser extent via single-top production channels (for recent reviews see \cite{JWK,DL}). Associated top production with a charged Higgs boson and top production via anomalous couplings are also of interest.

The QCD corrections to top production processes are typically quite large and need to be included for precise predictions. At current collider energies these corrections are dominated by soft-gluon emission. Thus one can derive excellent approximations to high-order corrections by calculating soft-gluon corrections. Such corrections need to be calculated through N$^3$LO to achieve percent-level or better precision.

In this paper we present high-precision theoretical results for top quark total and differential cross sections in various processes, including  top-antitop pair production \cite{NKprd90,NKprd91,NKprd91afb} in Section 2, single-top production \cite{NKsch,NKtW,NKtch,NKtchpt,NKsingletoppt} in Section 3, top production in association with a charged Higgs boson \cite{NKtH} in Section 4, and top production via anomalous couplings \cite{NKEM} in Section 5. 

\section{Top-antitop pair production}

The QCD corrections for $t{\bar t}$ production are quite large. Fixed-order NNLO calculations are now available based on a variety of analytical and numerical approaches by many groups (see \cite{DL,NKBP} for reviews). Soft-gluon corrections are dominant \cite{NKprd90,NKprd91,NKprd91afb} and they approximate the exact results very well at both NLO and NNLO.

We can calculate/resum these soft corrections at NNLL accuracy. Various approaches have been proposed over the years (see \cite{NKBP} for a review) and there are big differences in the accuracies of the various approximations. The method we use here has been the most successful in predicting the NNLO results, and the calculations of additional soft-gluon N$^3$LO corrections from NNLL resummation provide the best theoretical predictions for the production cross sections and differential distributions \cite{NKprd90,NKprd91,NKprd91afb}.

Approximate N$^3$LO (aN$^3$LO) predictions for the $t{\bar t}$ cross section are derived by adding the third-order soft-gluon corrections to the fixed-order results:

aN$^3$LO = LO + NLO + NNLO + soft-gluon N$^3$LO  corrections

\subsection{Top-antitop pair aN$^3$LO cross sections at the LHC and the Tevatron}

\begin{figure}
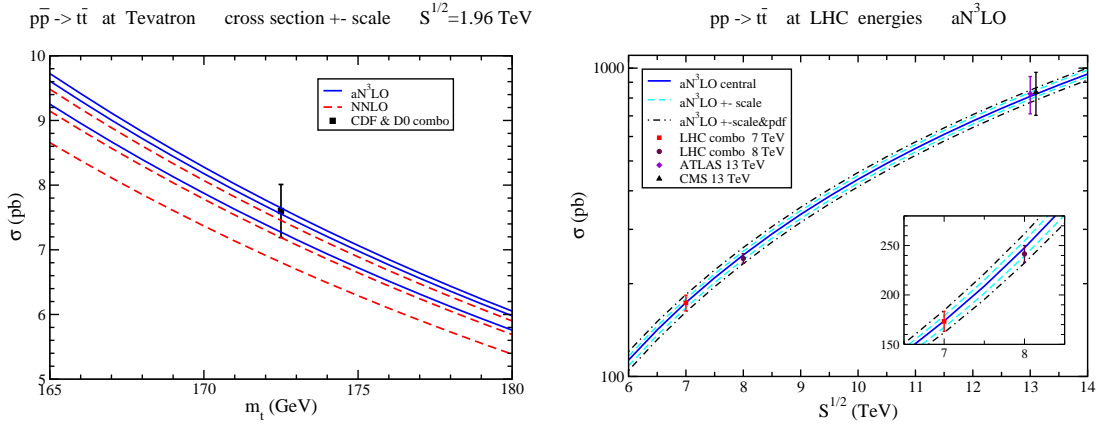

\begin{center}
\includegraphics[width=7cm]{tevatronaN3LOplot.eps}
\hspace{3mm}
\includegraphics[width=7cm]{ttSlhcaN3LOplot.eps}
\caption{Top-antitop aN$^3$LO cross sections at (left) the Tevatron compared with combined data from the CDF and D0 
collaborations at 1.96 TeV energy \cite{ttbartevcombo}; and (right) the LHC compared with combined 
data from ATLAS and CMS at 7 TeV \cite{tt7lhc} and 8 TeV \cite{tt8lhc}, and with the new data 
at 13 TeV from ATLAS \cite{ttATLAS13lhc} and CMS \cite{ttCMS13lhc}.}
\label{ttbartevlhcaN3LOplot}
\end{center}
\end{figure}

In Fig. \ref{ttbartevlhcaN3LOplot} we plot the aN$^3$LO \cite{NKprd90} 
$t{\bar t}$ total cross section with theoretical uncertainties at Tevatron and 
LHC energies and compare with recent data. 
The left plot shows results for the cross section at 1.96 Tevatron energy as 
a function of top quark mass while the plot on the right shows the cross section as a function of LHC energy. 
The agreement between data and the aN$^3$LO predictions is excellent in all cases.

The aN$^3$LO total $t{\bar t}$ cross section with $m_t=173.3$ GeV and MSTW2008 
NNLO pdf \cite{MSTW2008} takes the value of $810 {}^{+24}_{-16}{}^{+30}_{-32}$ pb at 13 LHC energy, where the first uncertainty is from scale variation and the second is the pdf uncertainty at 90\% CL.

\subsection{Relative size of the perturbative corrections}

The relative sizes of the higher-order corrections and the convergence of the perturbative 
series are of course of great interest. We write the perturbative series through third order 
in the strong coupling as
\beq
\sigma^{\rm aN^3LO}=\sigma^{(0)} \left[1+\frac{\sigma^{(1)}}{\sigma^{(0)}}
+\frac{\sigma^{(2)}}{\sigma^{(0)}}+\frac{\sigma^{({\rm a}3)}}{\sigma^{(0)}}\right] 
\eeq 
where $\sigma^{(0)}$ is the LO cross section, $\sigma^{(1)}$ and $\sigma^{(2)}$ are respectively 
the complete NLO and NNLO corrections, and $\sigma^{({\rm a}3)}$ are the aN$^3$LO corrections.

\begin{table}[htb]
\begin{center}
\begin{tabular}{|c|c|c|c|c|c|} \hline
corrections & Tevatron 1.96 TeV & LHC 7 TeV & LHC 8 TeV & LHC 13 TeV & LHC 14 TeV \\ \hline
$\sigma^{(1)}/\sigma^{(0)}$ & 0.236 & 0.470 & 0.476 & 0.493 & 0.496 \\ \hline 
$\sigma^{(2)}/\sigma^{(0)}$ & 0.106 & 0.178 & 0.177 & 0.172 & 0.170 \\ \hline 
$\sigma^{({\rm a}3)}/\sigma^{(0)}$ & 0.068 & 0.066 & 0.059 & 0.045 & 0.043 \\ \hline 
$\sigma^{\rm aN^3LO}/\sigma^{(0)}$ & 1.410 & 1.714 & 1.712 & 1.710 & 1.709 \\ \hline
\end{tabular}
\caption[]{The fractional contributions to the perturbative series for the 
$t{\bar t}$ cross section.}
\end{center}
\end{table}

In Table 1 we show the values of the ratios of the higher-order corrections to the LO 
cross section as well as the ratio  - in the last line - of the total aN$^3$LO cross section to 
the LO cross section. It is clear that higher-order corrections are very sizable for total cross sections, 
and this also holds for differential distributions. It is also obvious that NNLO is not enough! 
The aN$^3$LO corrections are clearly needed for precision physics. 

\subsection{Top-quark aN$^3$LO $p_T$ distributions at the LHC}

\begin{figure}
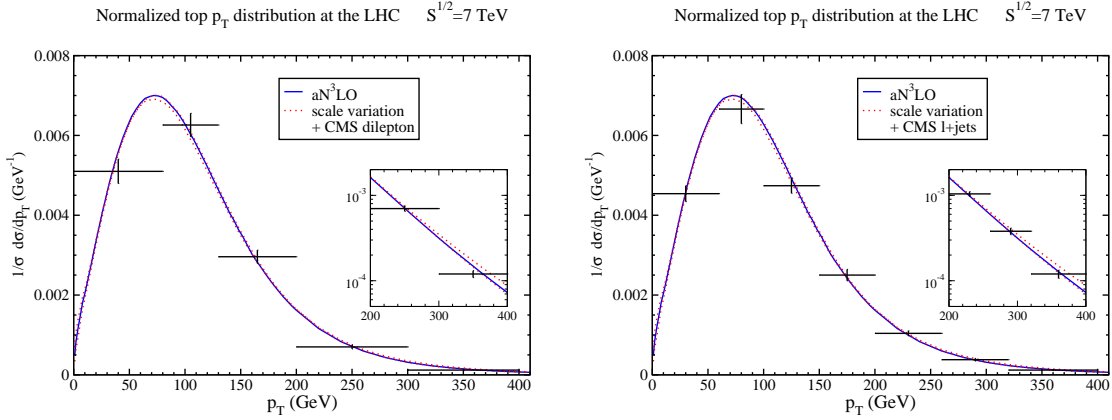

\begin{center}
\includegraphics[width=7cm]{pt7lhcnormCMSdileptplot.eps}
\hspace{5mm}
\includegraphics[width=7cm]{pt7lhcnormCMSleptjetplot.eps}
\caption{The aN$^3$LO normalized top-quark $p_T$ distributions at 7 TeV LHC energy compared with CMS dilepton (left) and lepton+jet (right) data \cite{CMStoppty7lhc}.}
\label{pt7lhcnormplot}
\end{center}
\end{figure} 

In Fig. \ref{pt7lhcnormplot} we plot the aN$^3$LO \cite{NKprd91} top-quark 
normalized transverse-momentum, $p_T$, distribution with scale variation at 
7 TeV LHC energy and compare with CMS data in the dilepton and lepton+jet 
channels, finding very good agreement in both cases. The inset
plots highlight the large-$p_T$ region. 

\begin{figure}
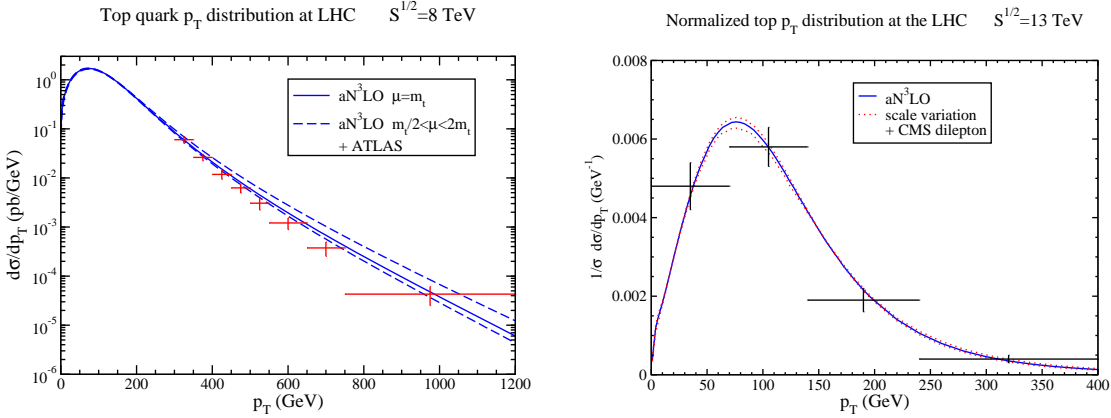

\begin{center}
\includegraphics[width=7cm]{pt8lhcaN3LOboostATLASplot.eps}
\hspace{5mm}
\includegraphics[width=7cm]{pt13lhcnormCMSdileptplot.eps}
\caption{The aN$^3$LO top-quark $p_T$ distribution (left) at the 8 TeV LHC compared with ATLAS boosted-top data \cite{ATLASboosttoppt8lhc}, and the normalized $p_T$ distribution (right) at the 13 TeV LHC compared with CMS \cite{CMStoppt13lhc} data.}
\label{toppt8-13lhc}
\end{center}
\end{figure}

In the left plot of Fig. \ref{toppt8-13lhc} we show the aN$^3$LO \cite{NKprd91} top-quark $p_T$ distribution at 8 TeV LHC energy and compare with boosted-top data from ATLAS; the right plot shows the normalized top-quark $p_T$ distribution at 13 TeV LHC energy compared with CMS data.

\subsection{Top-quark aN$^3$LO rapidity distributions at the LHC}

\begin{figure}
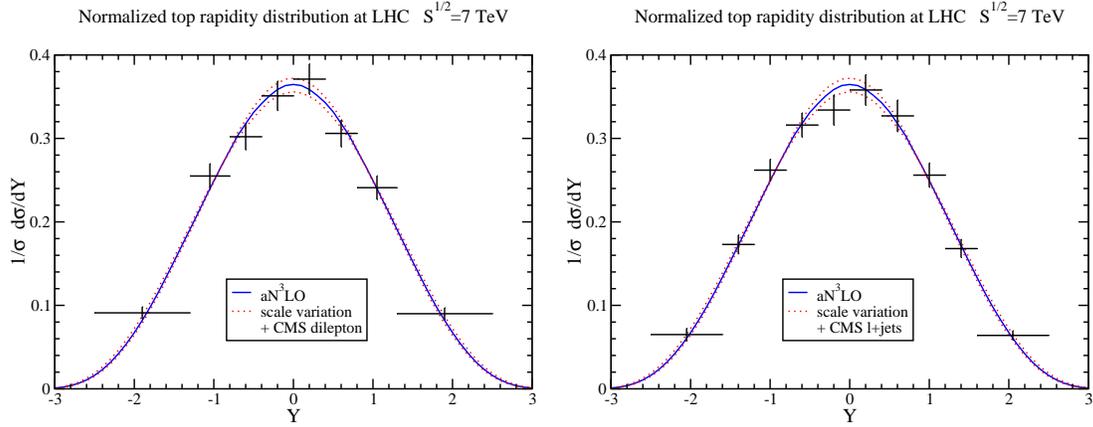

\begin{center}
\includegraphics[width=7cm]{y7lhcnormCMSdileptplot.eps}
\hspace{2mm}
\includegraphics[width=7cm]{y7lhcnormCMSleptjetplot.eps}
\caption{The aN$^3$LO normalized top-quark rapidity distributions at 7 TeV LHC energy compared with CMS dilepton (left) and lepton+jet (right) data \cite{CMStoppty7lhc}.}
\label{y7lhcnormplot}
\end{center}
\end{figure}

In Fig. \ref{y7lhcnormplot} we plot the aN$^3$LO \cite{NKprd91} top-quark normalized rapidity distribution at 7 TeV LHC energy and compare with CMS data in the dilepton and lepton+jet channels, finding excellent agreement between theory and data in both channels. 

\begin{figure}
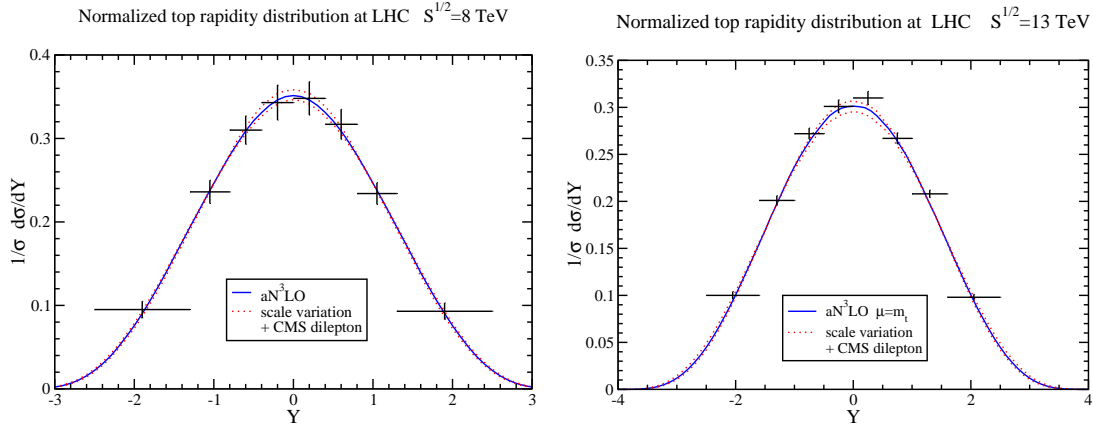

\begin{center}
\includegraphics[width=7cm]{y8lhcnormCMSdileptplot.eps}
\hspace{2mm}
\includegraphics[width=7cm]{ynorm13lhcCMSplot.eps}
\caption{The aN$^3$LO normalized top-quark rapidity distribution (left) at the 8 TeV LHC compared with CMS data \cite{CMStoppty8lhc} and (right) at the 13 TeV LHC compared with CMS \cite{CMStopy13lhc} data.}
\label{topy8-13lhc}
\end{center}
\end{figure}

In Fig. \ref{topy8-13lhc} we plot the aN$^3$LO \cite{NKprd91} top-quark normalized rapidity distribution at 8 TeV (left plot) 
and 13 TeV (right plot) LHC energies, and we compare with CMS data and find excellent agreement between theory and data at both energies. 

\subsection{Top-quark aN$^3$LO $p_T$ and rapidity distributions and $A_{\rm FB}$ at the Tevatron}

\begin{figure}
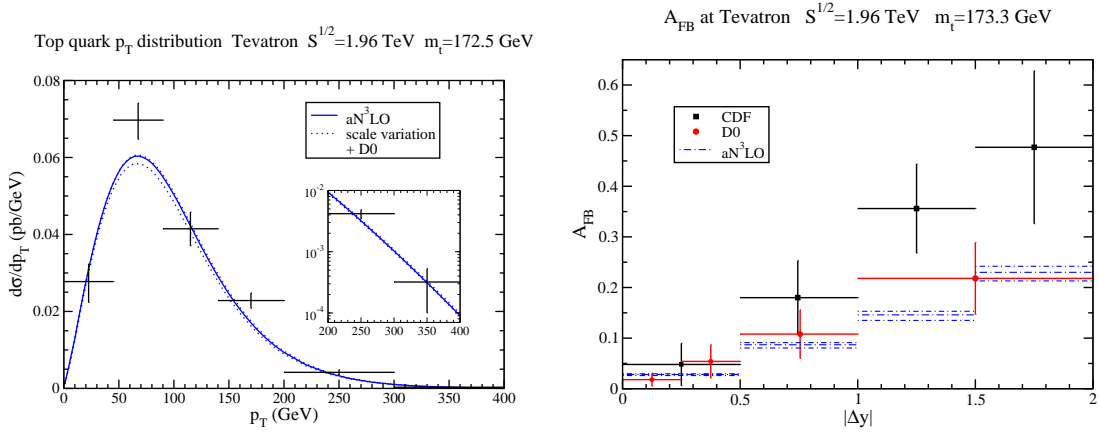

\begin{center}
\includegraphics[width=7cm]{pttevD0plot.eps}
\hspace{3mm}
\includegraphics[width=7cm]{AFBexpplot.eps}
\caption{The aN$^3$LO top-quark $p_T$ distribution at 1.96 TeV Tevatron energy compared with D0 \cite{D0pty} data (left); and the aN$^3$LO differential $A_{\rm FB}$ compared with CDF \cite{CDFafb} and D0 \cite{D0afb} data.}
\label{ptTevAFB}
\end{center}
\end{figure}

In the left plot of Fig. \ref{ptTevAFB} we plot the aN$^3$LO \cite{NKprd91}
top-quark $p_T$ distribution with scale variation at 1.96 TeV Tevatron 
energy and find very good agreement with D0 data.

The top forward-backward asymmetry at the Tevatron,
$A_{\rm FB} =[\sigma(y_t>0)-\sigma(y_t<0)]/[\sigma(y_t>0)+\sigma(y_t<0)]$, 
takes the aN$^3$LO \cite{NKprd91afb} QCD value -including EW corrections- 
of $10.0 \pm 0.6$\% in the $t{\bar t}$ frame.
We note that the corrections are large: the aN$^3$LO/NNLO ratio is 1.05. 

The top differential asymmetry 
$A^{\rm bin}_{\rm FB} =[\sigma^+_{\rm bin}(\Delta y)-\sigma^-_{\rm bin}(\Delta y)]/
[\sigma^+_{\rm bin}(\Delta y)+\sigma^-_{\rm bin}(\Delta y)]$, 
with $\Delta y=y_t-y_{\bar t}$, is shown at aN$^3$LO \cite{NKprd91afb} 
in the right plot of Fig. \ref{ptTevAFB} and compared to CDF and D0 data.

\section{Single-top production}

We continue with single-top production which can proceed through $t$-channel processes, $qb \rightarrow q' t$ and ${\bar q} b \rightarrow {\bar q}' t$; $s$-channel processes, $q{\bar q}' \rightarrow {\bar b} t$; and associated $tW$ production,  $bg \rightarrow tW^-$. 

\begin{figure}
\begin{center}
\includegraphics[width=9cm]{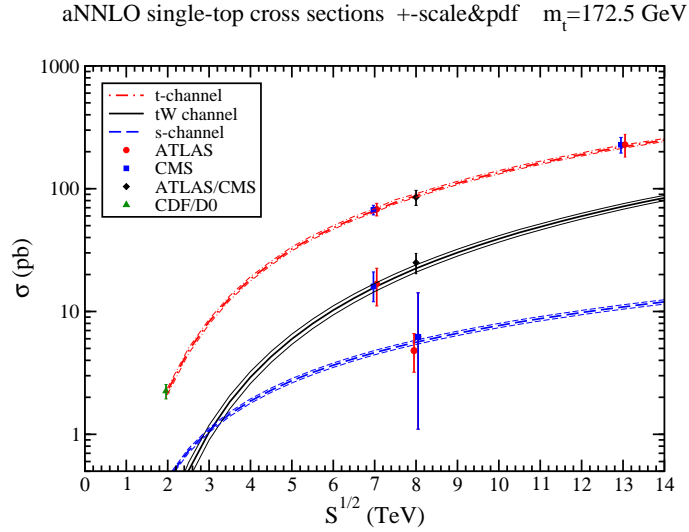}
\caption{Single-top aNNLO cross sections at the LHC compared to $t$-channel data from CDF/D0 combination \cite{tchcombotev} at 1.96 TeV, ATLAS \cite{ATLAStch7lhc} and CMS \cite{CMStch7lhc} at 7 TeV, ATLAS/CMS combination \cite{tchcombo8lhc} at 8 TeV, and ATLAS \cite{ATLAStch13lhc} and CMS \cite{CMStch13lhc} at 13 TeV; $s$-channel data from ATLAS \cite{ATLASsch8lhc} and CMS \cite{CMSsch8lhc} at 8 TeV;  and $tW$-channel data from ATLAS \cite{ATLAStW7lhc} and CMS \cite{CMStW7lhc} at 7 TeV, and ATLAS/CMS combination \cite{tWcombo8lhc} at 8 TeV.}
\label{singletopplot}
\end{center}
\end{figure}

The aNNLO single-top cross sections as functions of LHC energy are shown in Fig. \ref{singletopplot}. As can be clearly seen, there is excellent agreement of theory \cite{NKsch,NKtW,NKtch} with data for all three channels.

\subsection{Single-top $t$-channel production at aNNLO}

The aNNLO \cite{NKtch} single-top and single-antitop $t$-channel cross sections at the current 13 TeV LHC energy with $m_t=173.3$ GeV are, respectively, $136 {}^{+3}_{-1} \pm 3$ and $82 {}^{+2}_{-1} \pm 2$ pb. The errors indicated are from scale variation and pdf uncertainties from the MSTW2008 NNLO pdf at 90\% CL. 

\begin{figure}
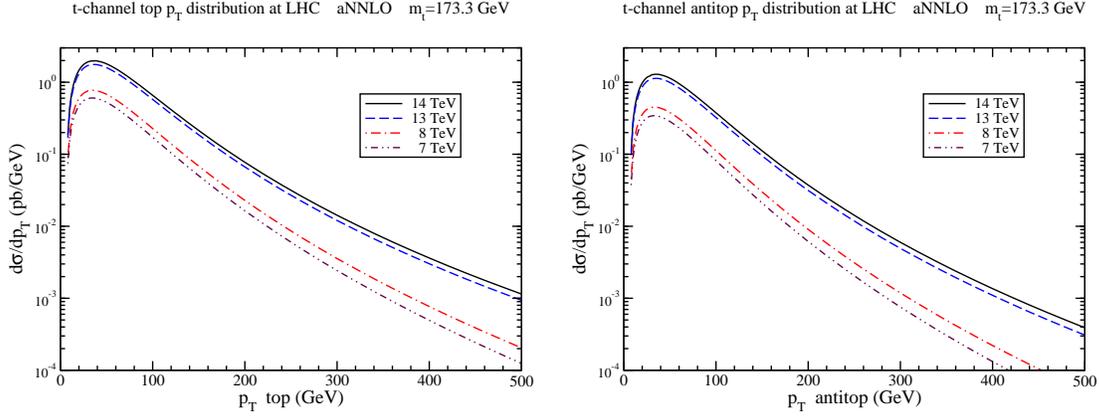

\begin{center}
\includegraphics[width=7cm]{pttoptchlhcplot.eps}
\hspace{3mm}
\includegraphics[width=7cm]{ptantitoptchlhcplot.eps} 
\caption{aNNLO $t$-channel top (left) and antitop (right) 
$p_T$ distributions at LHC energies.}
\label{pttoptch}
\end{center}
\end{figure}

The aNNLO \cite{NKtchpt,NKsingletoppt} top and antitop $p_T$ distributions in the $t$ channel at 7, 8, 13, and 14 TeV LHC energies are shown in Fig. \ref{pttoptch}.

\begin{figure}
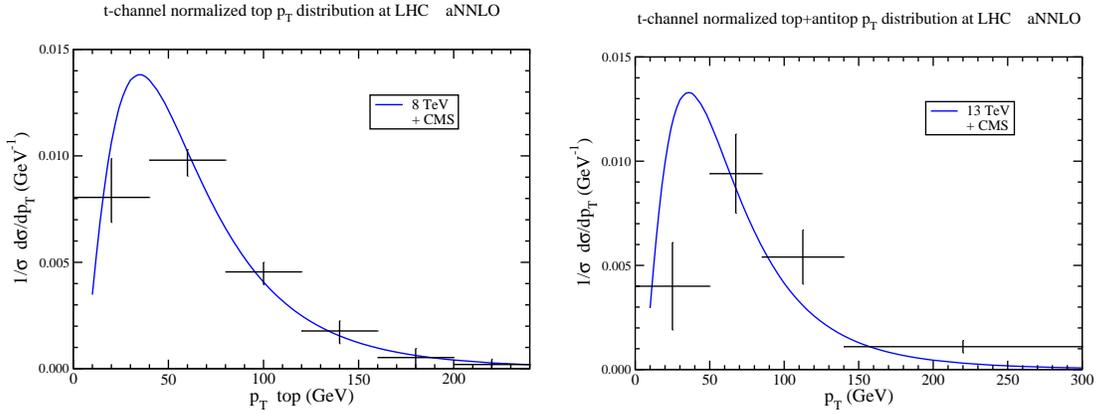

\begin{center}
\includegraphics[width=7cm]{ptnormtoptch8lhcCMSplot.eps}
\hspace{3mm}
\includegraphics[width=7cm]{ptnormtandtbartch13lhcCMSplot.eps} 
\caption{aNNLO $t$-channel normalized (left) top $p_T$ distribution compared with CMS data \cite{CMStchpt8lhc} at 8 TeV and (right) top+antitop $p_T$ distribution compared with CMS data \cite{CMStchpt13lhc} at 13 TeV.}
\label{ptnormtoptch}
\end{center}
\end{figure}

Some $t$-channel aNNLO normalized $p_T$ distributions at 8 and 13 TeV energies at the LHC are shown in Fig. \ref{ptnormtoptch} and compared with CMS data.

\subsection{Single-top $s$-channel production at aNNLO}

The aNNLO \cite{NKsch} single-top and single-antitop $s$-channel cross sections at the current 13 TeV LHC energy with $m_t=173.3$ GeV are, respectively, $7.07 \pm 0.13 {}^{+0.24}_{-0.22}$ and $4.10 \pm 0.05 {}^{+0.14}_{-0.16}$ pb.

\begin{figure}
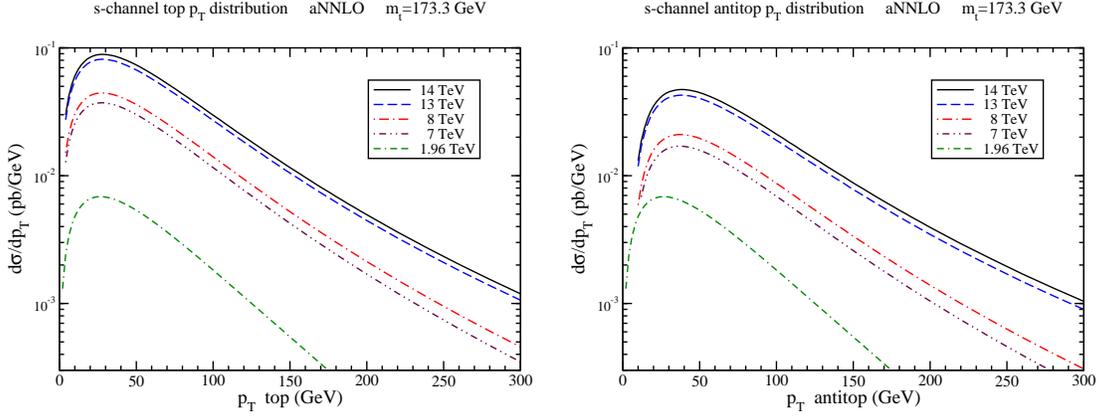

\begin{center}
\includegraphics[width=7cm]{pttopschplot.eps}
\hspace{3mm}
\includegraphics[width=7cm]{ptantitopschplot.eps} 
\caption{aNNLO $s$-channel top (left) and antitop (right) 
$p_T$ distributions at LHC and Tevatron energies.}
\label{pttopsch}
\end{center}
\end{figure}

The aNNLO \cite{NKsingletoppt} top and antitop $p_T$ distributions in the $s$ channel at 7, 8, 13, and 14 TeV LHC energies and at 1.96 TeV Tevatron energy are shown in Fig. \ref{pttopsch}.

\subsection{$tW^-$ production at aNNLO}

The aNNLO \cite{NKtW} $tW^-$ cross section at the current 13 TeV LHC energy with $m_t=173.3$ GeV is $35.2 \pm 0.9 {}^{+1.6}_{-1.7}$ pb. The cross section for ${\bar t}W^+$ is the same.

\begin{figure}
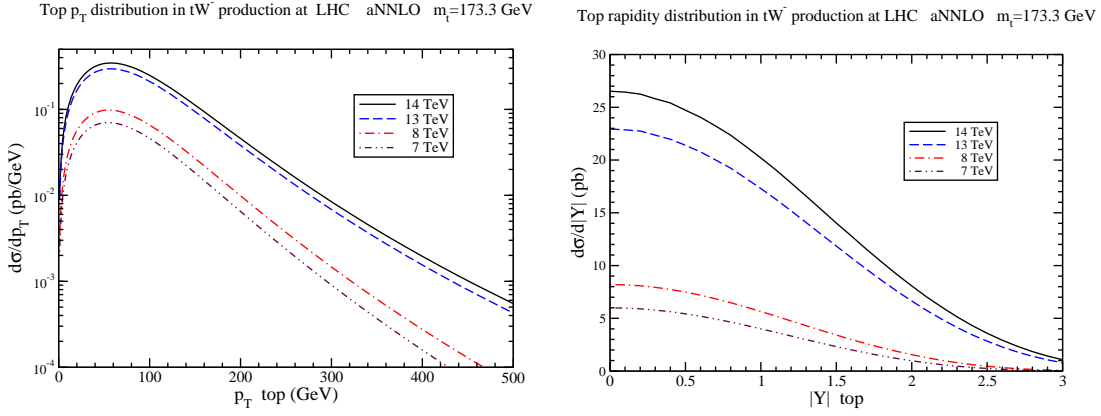

\begin{center}
\includegraphics[width=7cm]{pttoptWlhcplot.eps}
\hspace{3mm}
\includegraphics[width=7cm]{yabstopbgplot.eps} 
\caption{aNNLO top $p_T$ (left) and rapidity (right) 
distributions in $tW^-$ production at LHC energies.}
\label{ptytoptW}
\end{center}
\end{figure}

The aNNLO \cite{NKsingletoppt} top $p_T$ distributions in $tW^-$ production at LHC energies are shown in the left plot of Fig. \ref{ptytoptW} together with new results for the rapidity distributions in the right plot.

\section{$tH^-$ production at aNNLO}

We next present aNNLO \cite{NKtH} results for charged Higgs production in association with a top quark at LHC energies, using MMHT2014 NNLO pdf \cite{MMHT2014}.

\begin{figure}
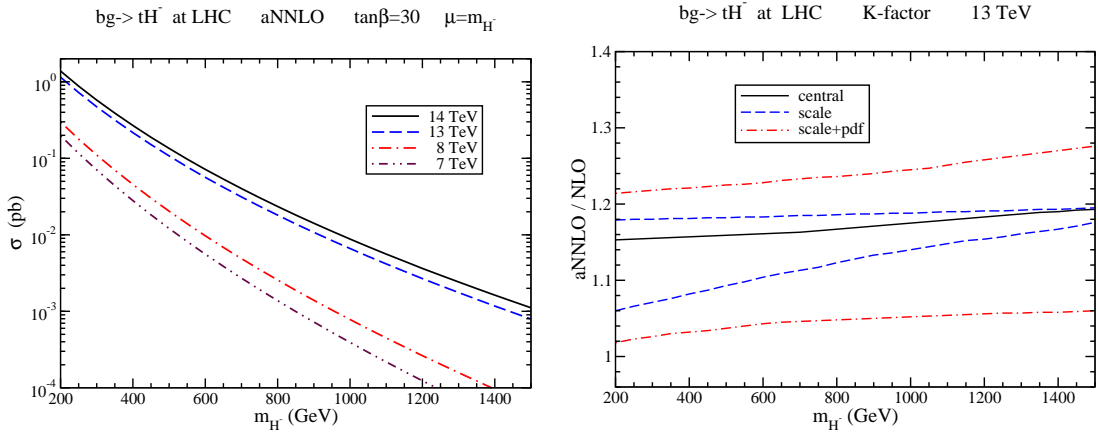

\begin{center}
\includegraphics[width=7cm]{chiggstn30plot.eps}
\hspace{3mm}
\includegraphics[width=7cm]{Kchiggs13lhcplot.eps} 
\caption{aNNLO cross sections at LHC energies (left) and 
$K$-factors at 13 TeV (right) for $tH^-$ production.}
\label{tH}
\end{center}
\end{figure}

In Fig. \ref{tH} we plot the aNNLO \cite{NKtH} cross sections at various LHC energies as functions of the charged Higgs mass, and also the associated $K$ factors at 13 TeV energy.

\begin{figure}
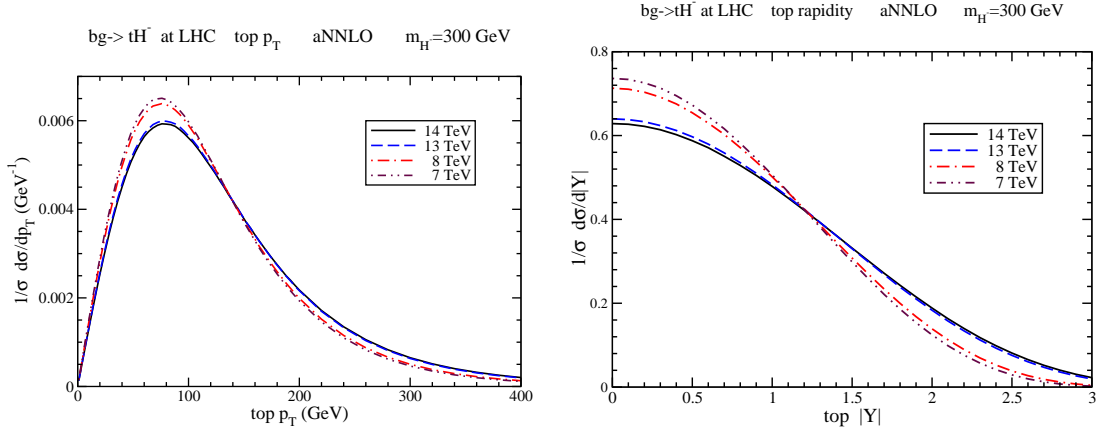

\begin{center}
\includegraphics[width=7cm]{ptnormtopchiggs300tn30plot.eps}
\hspace{3mm}
\includegraphics[width=7cm]{yabsnormtopchiggs300tn30plot.eps} 
\caption{aNNLO top $p_T$ (left) and rapidity (right) 
distributions in $tH^-$ production at LHC energies.}
\label{tHpty}
\end{center}
\end{figure}

In Fig. \ref{tHpty} we plot the aNNLO \cite{NKtH} normalized top-quark $p_T$ and rapidity distributions at LHC energies for a charged Higgs mass of 300 GeV.

\section{Top production via anomalous couplings}

Single-top production may proceed via anomalous couplings in theories beyond the Standard Model. Soft-gluon corrections for top production via anomalous gluon couplings were calculated through aNNLO in \cite{NKEM}. 
For processes $gu \rightarrow tg$ via anomalous $t$-$u$-$g$ couplings, 
the soft-gluon corrections are significant in enhancing the cross section and reducing the scale dependence.

\end{document}